\documentclass[sigconf]{acmart}
\AtBeginDocument{%
  }
\usepackage{multirow}
\usepackage{amsmath}
\usepackage{xspace}
\usepackage{graphicx}     % 插入图片
\usepackage{subcaption}

\usepackage{algorithm}
\usepackage{algorithmicx}
\usepackage{algpseudocode}
\usepackage{threeparttable}
\usepackage{url}

\makeatletter
\def\algbackskip{\hskip-\ALG@thistlm}
\makeatother

\setcopyright{acmlicensed}
\copyrightyear{2018}
\acmYear{2018}
\acmDOI{XXXXXXX.XXXXXXX}
\acmConference[Conference acronym 'XX]{Make sure to enter the correct
  conference title from your rights confirmation email}{June 03--05,
  2018}{Woodstock, NY}
\acmISBN{978-1-4503-XXXX-X/2018/06}

\newcommand{\pname}{{Co-PLMs}\xspace}

\begin{document}

\title{A Structure-Agnostic Co-Tuning Framework for LLMs and SLMs in Cloud-Edge Systems}

\author{Yuze Liu}
\affiliation{%
  \institution{Swinburne University of Technology}
  \city{Melbourne}
  \country{Australia}}
\email{yuzeliu@swin.edu.au}

\author{Yunhan Wang}
\affiliation{%
  \institution{Tongji University}
  \city{Shanghai}
  \country{China}}
\email{2354180@tongji.edu.cn}

\author{Tiehua Zhang}
\authornote{Corresponding author.}
\affiliation{%
  \institution{Tongji University}
  \city{Shanghai}
  \country{China}}
\email{tiehuaz@tongji.edu.cn}

\author{Zhishu Shen}
\affiliation{%
  \institution{Wuhan University of Technology}
  \city{Wuhan}
  \country{China}}
\email{z_shen@ieee.org}

\author{Cheng Peng}
\affiliation{%
  \institution{INFLY TECH (Shanghai) Co., Ltd.}
  \city{Shanghai}
  \country{China}}
\email{sophiapeng0426@hotmail.com}

\author{Libing Wu}
\affiliation{%
  \institution{Wuhan University}
  \city{Wuhan}
  \country{China}}
\email{wu@whu.edu.cn}

\author{Feng Xia}
\affiliation{%
  \institution{RMIT University}
  \city{Melbourne}
  \country{Australia}}
\email{feng.xia@rmit.edu.au}

\author{Jiong Jin}
\affiliation{%
  \institution{Swinburne University of Technology}
  \city{Melbourne}
  \country{Australia}}
\email{jiongjin@swin.edu.au}

\renewcommand{\shortauthors}{Trovato et al.}

\begin{abstract}
The surge in intelligent applications driven by large language models (LLMs) has made it increasingly difficult for bandwidth-limited cloud servers to process extensive LLM workloads in real time without compromising user data privacy. To solve these problems, recent research has focused on constructing cloud-edge consortia that integrate server-based LLM with small language models (SLMs) on mobile edge devices. Furthermore, designing collaborative training mechanisms within such consortia to enhance inference performance has emerged as a promising research direction. However, the cross-domain deployment of SLMs, coupled with structural heterogeneity in SLMs architectures, poses significant challenges to enhancing model performance. To this end, we propose Co-PLMs, a novel co-tuning framework for collaborative training of large and small language models, which integrates the process of structure-agnostic mutual learning to realize knowledge exchange between the heterogeneous language models. This framework employs distilled proxy models (DPMs) as bridges to enable collaborative training between the heterogeneous server-based LLM and on-device SLMs, while preserving the domain-specific insights of each device. The experimental results show that Co-PLMs outperforms state-of-the-art methods, achieving average increases of 5.38\% in Rouge-L and 4.88\% in EM. Our code has been released at https://github.com/papercode-DFL/Co-PLMs.
\end{abstract}






\received{20 February 2007}
\received[revised]{12 March 2009}
\received[accepted]{5 June 2009}

\maketitle

\section{Introduction}
The rapid development of large language models (LLMs) has sparked significant interest in their application to mobile devices and Web-of-Things (WoT), enabling intelligent services such as image recognition~\cite{ding2025gpt4image}, natural language understanding~\cite{liu2025grl}, and health monitoring~\cite{ma2025seek}, etc. LLMs are typically deployed on remote cluster servers to accommodate the significant computational resources required for inference~\cite{hao2024hybrid} and the need for ongoing updates via fine-tuning~\cite{yang2025fed}. However, the high volume of inference requests from a large number of edge devices creates communication bottlenecks, while the data demands of fine-tuning raise growing concerns about privacy leakage~\cite{nguyen2025device}. As a result, relying solely on server-based LLMs is neither efficient nor practical for real-world IoT and WoT applications~\cite{chen2025survey}.

To address these issues, recent research has focused on deploying small language models (SLMs), such as TinyLlama~\cite{zhang2024tinyllama}, Gemma~\cite{team2024gemma} and MobileVLM~\cite{chu2024mobilevlm}, on resource-constrained devices, and adopting federated learning (FL) to collaboratively train on-device SLMs for continual enhancement of model performance while preserving local data privacy~\cite{yang2025fed,cai2024edge,yu2024edge}. However, existing methods largely overlook the potential collaboration between SLMs and LLMs, failing to fully combine the strong generalization capabilities of LLMs with the efficiency of SLMs for on-device deployment. Therefore, designing an effective collaboration framework between on-device SLMs and server-based LLMs has emerged as a promising direction for future research~\cite{chen2025survey,deng2025crosslm,yan2024collaborate}.

Despite its potential, developing an effective collaboration framework for practical cloud-edge systems still faces two key challenges:
\begin{itemize}
    \item \textbf{\textit{Model Heterogeneity}}. In real-world scenarios, IoT and WoT devices exhibit diverse computational and communication capabilities due to varying hardware specification, necessitating the deployment of SLMs that differ in model size and model type~\cite{wang2023flexifed}. This disparity renders existing FL methods, such as FedLoRA~\cite{zhang2023fedpetuning} and FedAP~\cite{DBLP:journals/corr/abs-1902-00751}, which employ aggregation-based collaborative training, incapable of accommodating model heterogeneity~\cite{yi2023pfedlora}.
    \item \textbf{\textit{Domain Heterogeneity}}. Due to personalized usage patterns and diverse deployment environments, the local datasets on devices often exhibit domain-specific biases~\cite{shen2024tag}, which can lead to model performance degradation during the collaborative learning process~\cite{zhang2025fm2}.
\end{itemize}

To address these challenges, we propose a domain-aware heterogeneous co-tuning framework, termed \pname, which enables collaborative training between SLMs and LLMs within a cloud-edge system. This framework is composed of three key components, including initialization of the proxy model through model distillation from the server-based LLM, domain-specific tuning (DST) of the distilled proxy model (DPM) to capture domain biases within local datasets on devices and structure-agnostic mutual learning (SAML) between DPMs and language models to enable efficient bidirectional knowledge exchange. At the beginning of the collaborative learning process, the DPM is distilled from the server-based LLM and then distributed to each device, where domain adapters are integrated. During each collaborative training round, every device locally trains its DPM through DST to preserve domain-specific insights and performs SAML between the DPM with its local SLM to enable bidirectional knowledge transfer. Subsequently, the tunable parameters of the DPMs from all devices are uploaded to the server and aggregated to update the server-side DPM. The server then performs SAML between the LLM and the updated DPM to further update the LLM. Finally, the tunable parameters of the server-side DPM are redistributed to each device with the aim of propagating knowledge from the LLM.

The main advantages of our proposed framework are two folds: 1) DPMs act as bridges that enable collaborative training between heterogeneous language models within cloud-edge systems, where the exchange of DPM parameters facilitates effective knowledge sharing while preserving local data privacy. 2) Performing DST and SAML preserves domain-specific biases for each device while reducing computational costs through a parameter-efficient fine-tuning strategy. Furthermore, the communication overhead in \pname is significantly alleviated by exchanging only the tunable parameters of the DPM with the server during the collaborative learning process.

In summary, the main contributions of this work are:
\begin{itemize}
    \item We propose \pname, a novel co-tuning framework for collaborative training of large and small language models within cloud-edge system. In \pname, knowledge transfer between heterogeneous on-device SLMs and the server-based LLM relies on their respective distilled proxy models (DPMs), which act as bridges enabling bidirectional knowledge exchange between edge devices and the cloud server through structure-agnostic mutual learning (SAML). Meanwhile, domain-specific tuning (DST) of local DPMs helps preserve domain-specific insights during the collaborative learning process.
    \item We design structure-agnostic mutual learning (SAML), which incorporates bidirectional token alignment to address tokenizer mismatches and an logits pooling strategy to solve the problem divergence singularities during mutual learning, thereby enabling effective bidirectional knowledge transfer between heterogeneous language models.
    \item We conduct comprehensive experiments on the proposed framework using various publicly available language models across two multi-domain datasets, SNI and MMLU, along with a detailed analysis of communication overhead. Experimental results demonstrate that the proposed \pname significantly outperforms state-of-the-art baselines on both datasets while maintaining high communication efficiency throughout the collaborative learning process.
    \item We have released our source code publicly at \url{https://github.com/papercode-DFL/Co-PLMs} to facilitate further research and development in this field.
\end{itemize}
\section{Related  Work}
\subsection{LLMs Deployment at the Edge}
Traditional LLM deployment is based on centralized cloud servers, which provide sufficient computational and memory resources but suffer from network-dependent latency, high energy consumption, and privacy risks. In parallel, deploying full-scale LLMs directly on edge devices is also impractical, as their limited computation and memory capacity hinder real-time and high-quality inference. To address the aforementioned challenges, several recent studies have proposed collaborative frameworks for LLM deployment at the edge. Specifically, cloud-edge collaboration-based LLM frameworks, such as CE-CoLLM \cite{jin2025collm} and Edge-LLM \cite{cai2024edge}, offload part of the computation to the cloud while retaining latency-critical tasks at the edge through techniques like early exits, asynchronous context upload, and adaptive quantization. PerLLM \cite{yang2024perllm} further extends this direction by modeling inference scheduling as a multi-armed bandit problem under dynamic cloud-edge conditions. Edge-side optimization-based LLM frameworks, such as Edge-LLM \cite{yu2024edge} and LLMEdge \cite{ray2024llmedge}, directly address computation and memory bottlenecks in constrained devices. They exploit unified compression, adaptive layer tuning, and hardware-aware scheduling to improve efficiency, while incorporating quantization and lightweight APIs to support scalable and privacy-preserving deployment in IoT devices.

In summary, existing approaches can be broadly categorized into edge-side optimization techniques and collaborative frameworks between edge and cloud. These insights motivate the design of our framework, which introduces the SLM deployed at the edge, and further enhances lightweight, collaborative, and privacy-preserving deployment of LLMs at the edge.
\subsection{Federated Learning for LLMs}
In edge scenarios, clients are often constrained by limited computational costs and communication resources, in addition to stringent data privacy requirements. Federated learning for LLMs addresses these issues by enabling collaborative, cross-domain model adaptation, allowing clients to personalize models with their local data while keeping it private, and further reducing resource demands through parameter-efficient fine-tuning (PEFT). Recent works in PEFT-based federated learning, such as Federated Prompt Tuning\cite{zhao2023fedprompt}, FATE-LLM\cite{cheng2021secureboost}, and FedPETuning\cite{zhang2023fedpetuning}, leverage adapters, prompts and LoRA to minimize communication overhead efficiently. Notably, FedPFT~\cite{peng2024fedpft} constructs layer-wise sub-models with neuron-level alignment, FedLoRA\cite{wu2024fedlora} decomposes layers into shared and client-specific low-rank matrices with alternating training, and FedPETuning evaluates adapters, LoRA, and prefix tuning for lightweight updates under resource constraints. Furthermore, mutual knowledge transfer, as employed in FedCoLLM\cite{fan2024fedcollm} and FedMKT\cite{fan2024fedmkt}, introduces LLM deployed at the server and facilitates both server-sides LLMs and client-sides SLMs by utilizing adapters, knowledge distillation, and token alignment.

The aforementioned federated learning approaches either focus on efficient parameter updates or on mutual knowledge transfer between server and clients. Despite their effectiveness, challenges remain in further reducing communication overhead and supporting heterogeneous models with cross-domain setting, but these considerations are fully addressed in our proposed framework.
\section{Preliminaries}
\begin{figure}[t!]
    \centering
    \includegraphics[width=1\linewidth]{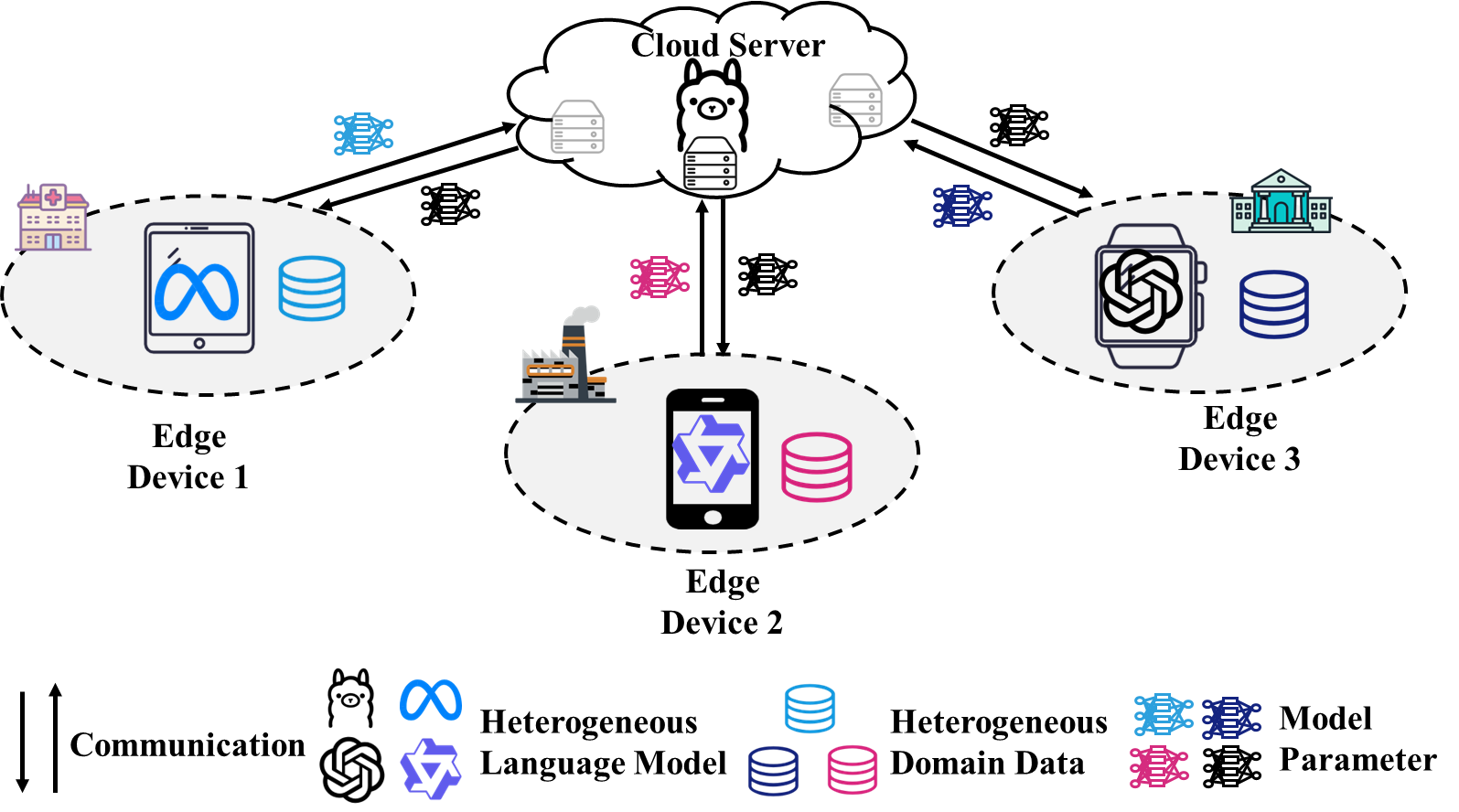}
  \caption{The architecture of the cloud-edge system.}
  \label{fig:1} 
\end{figure}
\noindent\textbf{Problem Definition}\quad
As illustrated in Figure~\ref{fig:1}, we consider an cloud-edge system consisting of a central server and a set of devices, denoted by $\mathcal{D} = \{d_i\}_{i=1}^{N}$, where $N$ represents the total number of devices. Each device hosts a SLM $m_i\in\mathcal{M}$ , while the server maintains a LLM denoted by $M$. Each SLM and the LLM possesses its own model structure, and the model–type mapping function is defined as $f_{mt}:\{M\}\cup\mathcal{M}\rightarrow\mathcal{T}$. $\mathcal{T}$ denotes the set of model structure types, encompassing various language model architectures such as GPT, Llama, and others. Each device $d_i$ and server owns a local private dataset denoted as $D^d_i$ and $D^s$, respectively. Specifically, local datasets among device are domain heterogeneity. Specifically, the local datasets across devices exhibit domain heterogeneity.

The server and devices collectively aim to improve the performance of the LLM and SLMs through collaborative learning while preserving data privacy. The collaborative learning process encompasses several rounds of training, in which each device uploads the locally trained model parameters to the server for  model aggregation. The updated model parameters are then distributed back to the devices. The objective function to be minimized for the entire collaborative learning process is defined as:
\begin{equation}
    \min_{\theta(M),\{\theta(m_{i})\}_{m_{i}\in\mathcal{M}}}\!\!\!\mathcal{L}(\theta(M), \theta(m_{1}), ..., \theta(m_{N});D^s, D^d_1, ..., D^d_N)
\end{equation}
where $\theta(M)$ and $\{\theta(m_{i})\}_{m_{i}\in\mathcal{M}}$ denote the trainable parameters of the server and the devices, respectively. Furthermore, we assume that the $N$ devices and the server perform the same question-answering (QA) task and employ heterogeneous model structure types, denoted by $|\mathcal{T}| > 1$. 

\noindent\textbf{Low-rank Adaptation}\quad
LoRA~\cite{hu2022lora} decomposes the update of the weight matrix $\emph{W}$ into the product of two low-rank matrices, under the assumption that fine-tuning updates are low-rank~\cite{aghajanyan2021intrinsic}. Here, $\emph{W}$ denotes the weight matrix of a linear layer in the model. For instance, in Transformer architectures, it may correspond to the query ($\emph{W}_q$), key ($\emph{W}_k$), value ($\emph{W}_v$) in the self-attention layer,. Specifically, LoRA has the following mathematical form:
\begin{equation}
    \emph{W}^{*} = \emph{W}_{0} + \triangle\emph{W} = \emph{W}_{0} + \emph{B}\emph{A}
\end{equation}
where $\emph{W}^{*},\emph{W}_{0}\in\mathcal{R}^{n\times m}$, $\emph{B}\in\mathcal{R}^{n\times r}$, and $\emph{A}\in\mathcal{R}^{r\times m}$, with $r\ll min(n,m)$. $\emph{W}_{0}$ denotes the pre-trained weight matrix, which remains frozen during the fine-tuning process, while $\mathbf{B}$ and $\mathbf{A}$ are trainable parameters. Owing to its training efficiency, LoRA has emerged as one of the most widely adopted parameter-efficient fine-tuning methods for training large language models~\cite{jiang2025dmlora}. The LoRA fine-tuning process for language models can be formulated as follows:
\begin{equation}
    \Tilde{\emph{m}} = f_{lora}(\phi_{lora}(\emph{m});\emph{m})
\end{equation}
where $\phi_{lora}(\emph{m}),\emph{m}\in\{M\}\cup\mathcal{M}$ denotes the trainable parameters during the LoRA fine-tuning process, and $\tilde{\emph{m}}$ represents the local language model after fine-tuning.

\section{Methodology}
% In this section, we present the proposed framework, \pname, a parameter-efficient collaborative learning method for cloud-edge systems deployed with heterogeneous LLM and SLMs, as illustrated in Figure~\ref{fig:overview}.

\begin{figure*}[t!]
    \centering
    \includegraphics[width=1\linewidth]{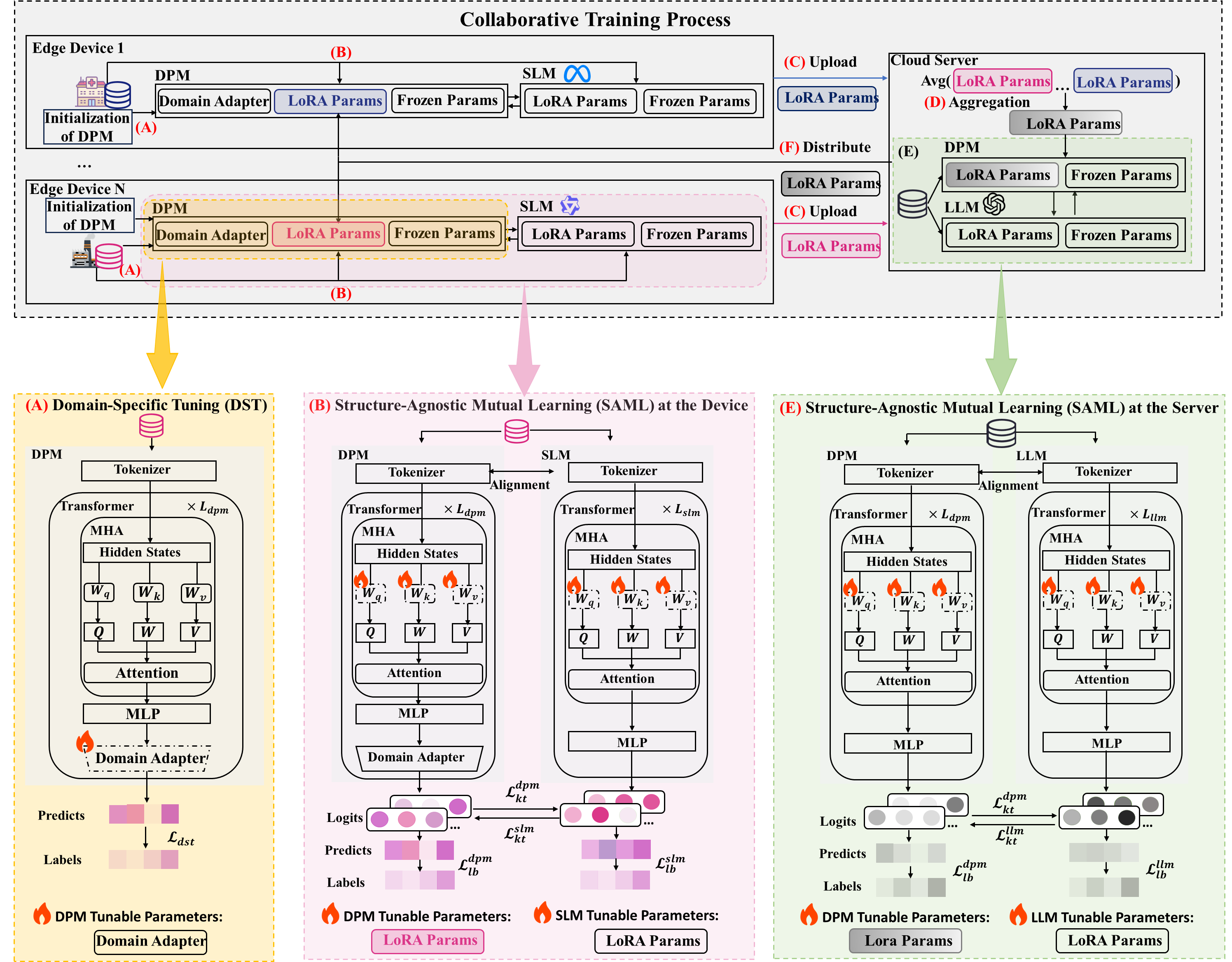}
  \caption{The overview of our proposed framework.}
  \label{fig:overview} 
\end{figure*}

An overview of the \pname architecture is presented in Figure~\ref{fig:overview}. In this section, we first describe the model distillation process from the LLM, which initializes the distilled proxy model (DPM). Next, domain-specific tuning (DST) is performed on the DPMs to capture domain biases, followed by structure-agnostic mutual learning (SAML), which enables bidirectional knowledge exchange between DPMs and language models on edge devices. On the server side, SAML is conducted between the updated DPM, aggregated from uploaded parameters, and the server-based LLM. Finally, the server distributes the parameters of its DPM to the devices to propagate knowledge throughout the system. 
\subsection{Initialization of Distilled Proxy Model}
Since the server-based LLM and the device-based SLMs have heterogeneous model structures, traditional FL approaches that directly aggregate local models for collaborative training are not applicable~\cite{wang2023flexifed}. Previous works have adopted proxy models that serve as bridges to facilitate knowledge transfer across models with heterogeneous architectures~\cite{peng2024fedpft,fan2024fedcollm}. However, existing approaches fail to account for the resource constraints of devices and to fully leverage the capabilities of the server-based LLM. To address these challenges, we adopt a knowledge distillation method, MiniLLM~\cite{gu2024minillm}, to generate a Transformer-based distilled proxy model (DPM) from the server-based LLM at the beginning of the collaborative learning process. The distillation process used to initialize the DPMs can be formulated as follows:
\begin{equation}
    \emph{m}^{p} = f_{kd}(\emph{M})
    \label{eq:init}
\end{equation}
where $\emph{m}^{p}$ denotes the DPM. After the distillation process is completed, $\emph{m}^{p}$ is distributed to each device to initialize the local DPMs, denoted as $\emph{m}^{p}_{i}, i \in [1, N]$.

\subsection{Domain-Specific Tuning}
To mitigate the performance degradation caused by domain heterogeneity in the local datasets of devices, we design a domain adapter for the DPM and propose a domain-specific tuning (DST) process to preserve its domain-dependent knowledge. We equip each Transformer layer of the DPM with a domain-aware adapter to extract domain-specific knowledge from the hidden representations of that layer. The domain adapter can be implemented using different neural architectures, such as a simple linear projection that maps the input to a lower-dimensional space or a multi-layer perceptron (MLP) that applies a sequence of nonlinear transformations. In this work, we employ a two-layer MLP with GeLU activation~\cite{hendrycks2016gaussian} as the domain adapter.

As illustrated in part (A) of Figure~\ref{fig:overview}, during the DST process, only the domain adapter is trained, while all other parameters in the DPM remain frozen. The training is conducted through supervised fine-tuning on local datasets, which can be formulated as follows:
\begin{equation}
    \min_{\phi_{da}(\emph{m}^{p}_{i})}\mathcal{L}_{dst}(\phi_{da}(\emph{m}^{p}_{i});D^{d}_{i})
\end{equation}
where $\mathcal{L}_{dst}(\cdot)$ denotes the loss function used in the DST process, and $\phi_{da}(\emph{m}^{p}_{i}), i\in[1,N]$ represents the tunable parameters in DPMs during DST. The process of DST can be defined as $f_{dst}(\emph{m}^{p}_i)$.
\subsection{Structure-Agnostic Mutual Learning}
To enable collaborative training under heterogeneous model architectures, we propose structure-agnostic mutual learning (SAML), which leverages DPMs as bridges to facilitate bidirectional knowledge transfer between the devices and the server. As illustrated in parts (B) and (C) of Figure~\ref{fig:overview}, SAML comprises pairs of a DPM and a language model that exchange knowledge by introducing a knowledge-transfer loss applied to their output logits.

\noindent\textbf{Bidirectional Token Alignment}\quad
A significant challenge in computing the mutual knowledge-transfer loss lies in the mismatch between the tokenizers of different language models. Take Qwen and Llama as an example. Consider the sentence “I utilize the map to travel.” When using the Qwen tokenizer, the sentence is segmented into the following tokens: [‘I’, ‘utilize’, ‘the’, ‘map’, ‘to’, ‘travel’]. In contrast, the Llama tokenizer segments it as [‘I’, ‘util’, ‘ize’, ‘the’, ‘map’, ‘to’, ‘travel’]. 

To address this issue, we employ a minimum edit distance algorithm with dynamic programming to construct token mappings between the two language models~\cite{fan2024fedmkt}. Using Qwen and Llama as illustrative examples, the token mapping aligns each Llama token with its closest corresponding Qwen token (e.g., ‘utilize’ for ‘util’). The bidirectional token alignment is formulated as $f_{dpm\rightarrow lm}(\emph{Y}^{p})$ and $f_{lm\rightarrow dpm}(\emph{Y}^{l})$, where $(\emph{Y}^{p})$ and $(\emph{Y}^{l})$ denote the output logits of the DPM and the language model, respectively.

\noindent\textbf{Output Logits Pooling for Knowledge Transfer Loss}\quad
Previous studies have pointed out that computing knowledge transfer loss typically relies on distribution-based methods such as Kullback–Leibler (KL) divergence~\cite{zhang2018deep,10.1145/3674729}. However, the large vocabulary size leads to sparse output distributions, which can cause divergence singularities~\cite{cover1999elements}. To solve this problem, we design an output logits pooling mechanism. Concretely, given the output logits $\emph{Y} = \{\emph{y}_{i}\}_{i = 1}^{S}$ with sequence length is $S$, each token vector $\emph{y}_{i}\in\mathcal{R}^{V}$ is pooled into a reduced representation $\tilde{\emph{y}}_{i}\in\mathcal{R}^{K+1}$ by preserving its Top-K largest components and aggregating the remaining ones into a single value. The pooling process can be formulated as follows:
\begin{equation}
    \tilde{\emph{y}} = f_{pool}(\emph{y})
\end{equation}
Furthermore, the knowledge transfer loss is defined as:
\begin{equation}
    \mathcal{L}_{kt}(\emph{Y},\emph{Y}^{'}) = \sum_{i = 1}^{S}KLD(\emph{y}_i,\emph{y}_i^{'}) 
\end{equation}
$S = \min(S_1,S_2)$, where $S_1$ and $S_2$ denote the sequence lengths of the output logits $\emph{Y}$ and $\emph{Y}^{'}$, respectively. During the process of SAML, the overall loss function $\mathcal{L}^{dpm}_{\phi_{lora}(\emph{m}^p)}$ for the DPM can be formulated as follows:
\begin{equation}
    \mathcal{L}^{dpm}_{\phi_{lora}(\emph{m}^p)} = \alpha\mathcal{L}_{kt}^{dpm}(f_{pool}(f_{lm\rightarrow dpm}(\emph{Y}^{l})),f_{pool}(\emph{Y}^{p})) + (1-\alpha)\mathcal{L}_{lb}^{dpm}
\end{equation}
where $\mathcal{L}_{lb}^{dpm}$ denotes the supervised finetuning loss. To achieve efficient fine-tuning in the cloud-edge system, SAML adopts a LoRA-based fine-tuning strategy. Similarly, the objective loss function $\mathcal{L}^{lm}_{\phi_{lora}(\emph{m})}$ for the corresponding language model in each input model pair within SAML can be computed as
\begin{equation}
    \mathcal{L}^{lm}_{\phi_{lora}(\emph{m})} = \beta\mathcal{L}_{kt}^{lm}(f_{pool}(f_{dpm\rightarrow lm}(\emph{Y}^{p})),f_{pool}(\emph{Y}^{l})) + (1-\beta)\mathcal{L}_{lb}^{lm}
\end{equation}
where $\emph{m}\in\{M\}\cup\mathcal{M}$. $\alpha$ and $\beta$ are the hyper-parameters that control the proportion of knowledge that from data or from the other model. The knowledge transfer between the DPMs and local language models is bidirectional, and both are trained on the private datasets in the device. The process of SAML can be defined as $f_{saml}(\emph{m}^{p},\emph{m})$.

\subsection{Overall Training Process}
The full collaborative training process of \pname is described in Algorithm~\ref{algo}. The workflow of \pname proceeds as follows:
\begin{enumerate}
    \item At the beginning of the collaborative learning process, the DPM is distilled from the server-based LLM. The server then distributes the DPM to each device, where domain adapters are subsequently integrated into it.
    \item In the t-th communication round, each device performs DST and SAML sequentially and then uploads the trainable LoRA parameters of its DPM to the server.
    \item The server aggregates the uploaded LoRA parameters by averaging them to update its DTM and then performs SAML. Afterwards, it distributes the updated LoRA parameters of the DTM to each device.
    \item Each device receives the LoRA parameters from the server and updates its local DPM accordingly.
\end{enumerate}
\renewcommand{\thealgorithm}{1}
\begin{algorithm}
\caption{\pname}\label{algo:1}
\begin{algorithmic}[1]
\Require The set of devices $\mathcal{D}$, the number of device $N$, the number of communication round $T$, the LLM in the server $M$, and the set of SLMs in the devices $\mathcal{M}$
\State Initialize the DPM $\emph{m}^{p}_s(1)$ at the server by Eq.\ref{eq:init}
\State Distribute the DPM to each device $\{\emph{m}_{i}^{p}(1)\}_{i=1}^{N}\leftarrow\emph{m}^{p}_s(1)$ and insert with domain adapters
\For{each round $t\in [1,2,...,T]$}
\State // Device side.
\For{each device $d_i\in\mathcal{D}$ (in parallel)}
\State$\Bar{\emph{m}}_{i}^{p}(t)\leftarrow f_{dst}(\emph{m}_{i}^p(t))$\Comment{Domain-specific tuning}
\State Token alignment between $\emph{m}_{i}^{p}(t)$ and $\emph{m}_{i}(t)$ 
\State $\hat{\emph{m}}^{p}_i(t),\emph{m}_i(t+1),\phi_{lora}(\hat{\emph{m}}^{p}_i(t))\leftarrow f_{saml}(\Bar{\emph{m}}_{i}^{p}(t), \emph{m}_i(t))$ \Comment{Structure-agnostic mutual learning at the device}
\State upload $\phi_{lora}(\hat{\emph{m}}^{p}_i(t))$ to the server
\EndFor
\State // Server side.
\State $\phi_{lora}(\emph{m}^{p}_s(t)) = \frac{1}{N}\sum_{i=1}^{N}\phi_{lora}(\hat{\emph{m}}^{p}_i(t))$ \Comment{Model parameters aggregation}
\State Update $\emph{m}^{p}_s(t)$ by $\phi_{lora}(\emph{m}^{p}_s(t))$
\State  $\emph{m}^{p}_s(t+1),\emph{M}(t+1),\phi_{lora}(\emph{m}^{p}_s(t+1))\leftarrow f_{saml}(\emph{m}_{s}^{p}(t), \emph{M}(t))$ \Comment{Structure-agnostic mutual learning at the server}
\State Distribute $\phi_{lora}(\emph{m}^{p}_s(t+1))$ to each device
\State // Device side.
\For{each device $d_i\in\mathcal{D}$ (in parallel)}
\State $\phi_{lora}(\emph{m}^{p}_i(t+1))\leftarrow\phi_{lora}(\emph{m}^{p}_s(t+1))$
\State Update $\emph{m}^{p}_i(t+1)$ by $\phi_{lora}(\emph{m}^{p}_i(t+1))$
\EndFor
\EndFor
\end{algorithmic}
\label{algo}
\end{algorithm}

\section{Experiment}
\subsection{Experiment Setup}
We assume an cloud-edge system comprising one cloud server and $N = 3$ edge devices to evaluate \pname using various publicly available language models. The details are as below:

\noindent\textbf{Models}\quad
% \subsubsection{Models}
To evaluate \pname, we consider a setup comprising, GPT-J-6B \cite{wang2021gpt}, deployed on the cloud server, and three SLMs, Bloom-1.1B \cite{workshop2022bloom}, Llama 2-1.3B \cite{xia2023sheared}, and Qwen2.5-1.5B \cite{yang2025qwen2}, deployed on the edge devices.

\noindent\textbf{Datasets}\quad 
We evaluate the performance of our framework on two multi-domain question-answering (QA) datasets, SNI \cite{wang2022super} and MMLU \cite{hendrycksmeasuring}. SNI encompasses data from 33 domains, with each sample accompanied by a corresponding instruction. MMLU is organized as a multiple-choice QA dataset spanning 57 domains.

\noindent\textbf{Data Partition}\quad 
In this experiment, the local datasets are sampled from a Dirichlet distribution $Dir(\lambda)$ to emulate domain heterogeneity among devices~\cite{liu2025towards}. The parameter $\lambda$ controls the degree of data domain skewness (DDS). Specifically, as $\lambda\rightarrow 0$, the local datasets on each device become increasingly biased toward a single domain. The server’s local dataset is uniformly sampled from the global dataset. Each local dataset contains 1,000 samples, of which 80\% are used for training on both the devices and the server~\cite{zhang2023adaptive}.

\noindent\textbf{Baselines}\quad 
We conduct a comparative analysis of \pname against the following baselines:
\begin{itemize}
\item \textbf{Standalone} allows both edge devices and the cloud server to independently fine-tune their respective language models using local datasets, without involving any form of collaborative training.
\item \textbf{FedLoRA}~\cite{zhang2023fedpetuning} applies low-rank adaptation for parameter-efficient fine-tuning, with the rank $r$ set to 8. All edge devices deploy language models with the same architecture and communicate with the cloud server solely by transmitting the updated low-rank adaptation matrices. The server aggregates the received models using simple averaging.
\item \textbf{FedAP}~\cite{DBLP:journals/corr/abs-1902-00751} inserts adapters into both the multi-head attention module and the feed-forward network within each Transformer layer of the language models deployed on edge devices. During the training process, the edge devices update only the adapters while keeping the main model frozen. The updated adapters are then uploaded to the cloud server, where model aggregation is performed through averaging.
\item \textbf{FedCoLLM}~\cite{fan2024fedcollm} introduces lightweight LoRA modules for mutual knowledge transfer between server-based LLM and on-device SLMs, and employs secure aggregation to preserve privacy during the transfer process.
\item \textbf{FedMKT}~\cite{fan2024fedmkt} employs a federated mutual knowledge transfer framework that enables selective distillation between the server-side LLM and device-side SLMs, wherein edge devices transmit their output logits and training losses to facilitate bidirectional knowledge exchange.
\end{itemize}
\vspace{-0.2em} 

\noindent\textbf{Evaluation Metrics}\quad
For the QA task, we evaluate the performance of language models using Rouge-L and Exact Match (EM). 
\begin{itemize}
\item \textbf{Rouge-L} measures the overlap of the longest common subsequence between the generated and reference answers, effectively capturing both content similarity and fluency.
\item \textbf{Exact Match} calculates the percentage of generated answers that exactly match the reference answers, reflecting the model’s ability to produce fully correct responses.
\end{itemize}
% \vspace{-1em} 

\subsection{Performance Analysis}
% 第一段介绍 homo 和 hetergeneous的setting
% 第二/三段我们的方法在heterogeneous的时候平均比其他的多多少 homo的时候平均多多少
% 因为domain差异变大导致的和standlone相比的performance的降低
We simulate varying levels of domain heterogeneity in the cloud-edge system by setting different values of $\lambda \in \{0.1, 1\}$ for DDS. All experiments are conducted five times with different random seeds, and the average results are reported in Table~\ref{tab:1}. We analyze \pname under two experimental settings: homogeneous and heterogeneous model structures on the devices. Firstly, we can observe that our framework consistently outperforms the Standalone approach under both homogeneous and heterogeneous model structure settings on the devices, achieving performance gains of 0.30\% to 12.50\% in Rouge-L and 1.40\% to 13.10\% in EM across different language models. This indicates that collaborative learning enhances overall model performance, even under scenarios with varying levels of domain heterogeneity.

For the setting of homogeneous model structure on the devices, methods are evaluated under the configuration where edge devices are deployed with the same SLM architecture, and the performance results from identical devices are reported in Table~\ref{tab:1}. We conduct experiments using three representative language models, Bloom-1.1B, LLaMA2-1.3B, and Qwen2.5-1.5B, across FedLoRA, FedAP, and \pname. It is worth noting that FedLoRA and FedAP are not applicable for evaluating server performance, as LLMs cannot be deployed on the server under their experimental settings. In contrast, \pname deploys GPT-J-6B on the server for evaluation. As shown in Table~\ref{tab:1}, \pname consistently outperforms FedLoRA and FedAP across different deployed language models, with performance gains of 0.40\% to 26.10\% in Rouge-L and 0.20\% to 17.30\% in EM. On average, \pname achieves gains in 2.38\% , 2.55\% and 2.83\% in Rouge-L and 2.35\% , 2.45\% and 2.68\% in EM compared to FedLoRA on Bloom-1.1B, LLaMA2-1.3B and Qwen2.5-1.5B, respectively. Similarly, \pname demonstrates average improvements of 5.90\% , 18.78\% and 5.53\% in Rouge-L and 4.98\% , 12.15\% and 5.10\% in EM, compared to FedAP.

To evaluate the performance of \pname under the setting of heterogeneous model structures on the devices, we configure the experimental setup such that the edge devices are deployed with different SLM architectures. Specifically, Device-1 is deployed with Bloom-1.1B, Device-2 with LLaMA2-1.3B, Device-3 with Qwen2.5-1.5B and server with GPT-J-6B. As presented in Table~\ref{tab:1}, \pname significantly outperforms FedCoLLM and FedMKT across all edge devices and the cloud server, achieving performance gains ranging from 0.40\% to 12.80\%. Specifically, 
\pname achieve performance gains of 0.70\% to 11.90\% and 0.40\% to 9.50\% in Rouge-L and 1.60\% to 12.80\% and 0.90\% to 7.40\% in EM compared to FedCoLLM and FedMKT. On average, \pname achieves an improvement of 5.48\%, 3.45\% and 7.05\% in Rouge-L and 5.50\%, 4.53\% and 9.93\% in EM compared to FedCoLLM on Device-1,  Device-2 and Device-3, respectively. Similarly, \pname demonstrates average improvements of 3.53\%, 2.58\% and 4.50\% in Rouge-L and 3.43\%, 2.85\% and 5.60\% in EM compared with FedMKT.

It's worth noting that the performance across all methods except Standalone under the setting of $\lambda = 0.1$ is lower than that under $\lambda = 1$, with an average drop of 7.40\% in Rouge-L and 3.90\% in EM. This decline is due to the domain distribution of the datasets, which increases heterogeneity and makes it more challenging for the models to learn effectively. Specifically, on the SNI dataset, when $\lambda = 1$, \pname under the setting of heterogeneous model structures on the devices obtains average improvements over Standalone, by 1.97\%. However, when $\lambda = 0.1$, the average of improvement is reduced to 1.59\%, indicating the sensitivity of collaborative training to domain heterogeneity.
\begin{table*}[]
\caption{Summary of the comparison results of \pname on the SNI and MMLU datasets. The best results are highlighted in bold.}
\label{tab:1}
\begin{threeparttable}
\resizebox{0.95\linewidth}{!}{
% \usepackage{multirow}
\begin{tabular}{cccc|cc|cc|cc|cc}
\hline
\multirow{2}{*}{Dataset}                    & \multirow{2}{*}{DDS}                                  & \multirow{2}{*}{\begin{tabular}[c]{@{}c@{}}Model Structures\\ on the Devices\end{tabular}} & \multirow{2}{*}{Method} & \multicolumn{2}{c|}{\begin{tabular}[c]{@{}c@{}}Device-1\\ (Bloom-1.1B)\end{tabular}} & \multicolumn{2}{c|}{\begin{tabular}[c]{@{}c@{}}Device-2\\ (Llama2-1.3B)\end{tabular}} & \multicolumn{2}{c|}{\begin{tabular}[c]{@{}c@{}}Device-3\\ (Qwen2.5-1.5B)\end{tabular}} & \multicolumn{2}{c}{\begin{tabular}[c]{@{}c@{}}Server\\ (GPT-J-6B)\end{tabular}} \\ \cline{5-12} 
                                            &                                                       &                                                                                      &                         & Rouge-L                                   & EM                                       & Rouge-L                                   & EM                                        & Rouge-L                                    & EM                                        & Rouge-L                                & EM                                     \\ \hline
\multicolumn{1}{c|}{\multirow{14}{*}{SNI}}  & \multicolumn{1}{c|}{\multirow{7}{*}{$\lambda = 0.1$}} & \multicolumn{1}{c|}{-}                                                               & Standalone              & 39.9                                      & 29.6                                     & 38.7                                      & 24.1                                      & 43.7                                       & 31.2                                      & 53.5                                   & 37.3                                   \\ \cline{3-4}
\multicolumn{1}{c|}{}                       & \multicolumn{1}{c|}{}                                 & \multicolumn{1}{c|}{\multirow{3}{*}{Homogeneous}}                                    & FedLoRA                 & 33.7                                      & 23.3                                     & 40.4                                      & 28.2                                      & 44.7                                       & 29.2                                      & -                                      & -                                      \\
\multicolumn{1}{c|}{}                       & \multicolumn{1}{c|}{}                                 & \multicolumn{1}{c|}{}                                                                & FedAP                   & 28.2                                          & 22.1                                         & 16.3                                          & 12.1                                          &39.8                                            &28.9                                           & -                                      & -                                      \\
\multicolumn{1}{c|}{}                       & \multicolumn{1}{c|}{}                                 & \multicolumn{1}{c|}{}                                                                & Ours                    & 36.8                                      & 24.5                                     & 40.8                                      & 29.4                                      & \textbf{54.8}                              & 37.8                                      & 58.9                                   & 37.8                                   \\ \cline{3-4}
\multicolumn{1}{c|}{}                       & \multicolumn{1}{c|}{}                                 & \multicolumn{1}{c|}{\multirow{3}{*}{Heterogeneous}}                                  & FedCoLLM                & 32.6                                      & 21.7                                     & 38.2                                      & 26.4                                      & 48.7                                       & 29.8                                      & 51.4                                   & 37.2                                   \\
\multicolumn{1}{c|}{}                       & \multicolumn{1}{c|}{}                                 & \multicolumn{1}{c|}{}                                                                & FedMKT                  & 36.7                                      & 24.3                                     & 41.4                                      & 27.8                                      & 52.8                                       & 34.3                                      & 54.7                                   & 35.8                                   \\
\multicolumn{1}{c|}{}                       & \multicolumn{1}{c|}{}                                 & \multicolumn{1}{c|}{}                                                                & Ours                    & \textbf{42.5}                             & \textbf{31.7}                            & \textbf{41.8}                             & \textbf{30.2}                             & 54.3                                       & \textbf{39.2}                             & \textbf{59.2}                          & \textbf{39.5}                          \\ \cline{2-12} 
\multicolumn{1}{c|}{}                       & \multicolumn{1}{c|}{\multirow{7}{*}{$\lambda = 1$}}   & \multicolumn{1}{c|}{-}                                                               & Standalone              & 41.2                                      & 27.0                                     & 44.6                                      & 23.5                                      & 38.2                                       & 21.9                                      & 58.6                                   & 34.1                                   \\ \cline{3-4}
\multicolumn{1}{c|}{}                       & \multicolumn{1}{c|}{}                                 & \multicolumn{1}{c|}{\multirow{3}{*}{Homogeneous}}                                    & FedLoRA                 & 37.9                                      & 24.6                                     & 48.9                                      & 28.1                                      & 45.8                                       & 29.6                                      & -                                      & -                                      \\
\multicolumn{1}{c|}{}                       & \multicolumn{1}{c|}{}                                 & \multicolumn{1}{c|}{}                                                                & FedAp                   & 33.6                                          & 20.3                                         &  23.3                                         & 14.4                                          & 44.6                                           & 26.8                                          & -                                      & -                                      \\
\multicolumn{1}{c|}{}                       & \multicolumn{1}{c|}{}                                 & \multicolumn{1}{c|}{}                                                                & Ours                    & 39.1                                      & 24.8                                     & 49.4                                      & 28.1                                      & 47.1                                       & 29.9                                      & 57.6                                   & 33.8                                   \\ \cline{3-4}
\multicolumn{1}{c|}{}                       & \multicolumn{1}{c|}{}                                 & \multicolumn{1}{c|}{\multirow{3}{*}{Heterogeneous}}                                  & FedCoLLM                & 38.4                                      & 21.8                                     & 46.4                                      & 24.8                                      & 38.8                                       & 23.4                                      & 54.2                                   & 32.9                                   \\
\multicolumn{1}{c|}{}                       & \multicolumn{1}{c|}{}                                 & \multicolumn{1}{c|}{}                                                                & FedMKT                  & 39.2                                      & 27.1                                     & 47.6                                      & 24.3                                      & 41.2                                       & 26.8                                      & 56.1                                   & 32.8                                   \\
\multicolumn{1}{c|}{}                       & \multicolumn{1}{c|}{}                                 & \multicolumn{1}{c|}{}                                                                & Ours                    & \textbf{44.9}                             & \textbf{30.2}                            & \textbf{51.5}                             & \textbf{29.7}                             & \textbf{50.7}                              & \textbf{34.2}                             & \textbf{59.7}                          & \textbf{37.8}                          \\ \hline
\multicolumn{1}{c|}{\multirow{14}{*}{MMLU}} & \multicolumn{1}{c|}{\multirow{7}{*}{$\lambda = 0.1$}} & \multicolumn{1}{c|}{-}                                                               & Standalone              & 44.0                                      & 31.5                                     & 51.6                                      & 23.5                                      & 49.7                                       & 33.5                                      & 56.8                                   & 36.6                                   \\ \cline{3-4}
\multicolumn{1}{c|}{}                       & \multicolumn{1}{c|}{}                                 & \multicolumn{1}{c|}{\multirow{3}{*}{Homogeneous}}                                    & FedLoRA                 &41.2                                           & 27.3                                         & 45.6                                      & 20.0                                      & 54.2                                       & 35.5                                      & -                                      & -                                      \\
\multicolumn{1}{c|}{}                       & \multicolumn{1}{c|}{}                                 & \multicolumn{1}{c|}{}                                                                & FedAp                   &39.1                                           & 24.5                                         & 29.8                                         & 19.2                                          & 49.1                                           & 34.0                                          & -                                      & -                                      \\
\multicolumn{1}{c|}{}                       & \multicolumn{1}{c|}{}                                 & \multicolumn{1}{c|}{}                                                                & Ours                    & 44.0                                      & 31.8                                     & 52.1                                      & 27.1                                      & 50.7                                       & 36.1                                      & 55.7                                   & 34.8                                   \\ \cline{3-4}
\multicolumn{1}{c|}{}                       & \multicolumn{1}{c|}{}                                 & \multicolumn{1}{c|}{\multirow{3}{*}{Heterogeneous}}                                  & FedCoLLM                & 41.2                                      & 29.4                                     & 48.5                                      & 21.6                                      & 51.1                                       & 32.7                                      & 51.8                                   & 34.2                                   \\
\multicolumn{1}{c|}{}                       & \multicolumn{1}{c|}{}                                 & \multicolumn{1}{c|}{}                                                                & FedMKT                  & 42.8                                      & 30.1                                     & 48.3                                      & 27.2                                      & 53.7                                       & 34.5                                      & 55.7                                   & 36.9                                   \\
\multicolumn{1}{c|}{}                       & \multicolumn{1}{c|}{}                                 & \multicolumn{1}{c|}{}                                                                & Ours                    & \textbf{44.3}                             & \textbf{31.0}                            & \textbf{52.9}                             & \textbf{28.5}                             & \textbf{57.2}                              & \textbf{39.4}                             & \textbf{60.4}                          & \textbf{38.0}                          \\ \cline{2-12} 
\multicolumn{1}{c|}{}                       & \multicolumn{1}{c|}{\multirow{7}{*}{$\lambda = 1$}}   & \multicolumn{1}{c|}{-}                                                               & Standalone              & 43.8                                      & 24.8                                     & 40.1                                      & 23.5                                      & 41.4                                       & 20.6                                      & 52.7                                   & 29.7                                   \\ \cline{3-4}
\multicolumn{1}{c|}{}                       & \multicolumn{1}{c|}{}                                 & \multicolumn{1}{c|}{\multirow{3}{*}{Homogeneous}}                                    & FedLoRA                 & 42.3                                          &22.8                                          & 41.5                                      & 26.0                                      & 47.8                                       & 26.5                                      & -                                      & -                                      \\
\multicolumn{1}{c|}{}                       & \multicolumn{1}{c|}{}                                 & \multicolumn{1}{c|}{}                                                                & FedAp                   &40.1                                           &20.6                                          & 42.1                                          & 17.8                                          & 48.2                                           &21.6                                           & -                                      & -                                      \\
\multicolumn{1}{c|}{}                       & \multicolumn{1}{c|}{}                                 & \multicolumn{1}{c|}{}                                                                & Ours                    & 44.7                                      & 26.3                                     & 44.3                                      & 27.5                                      & 51.2                                       & 27.7                                      & 58.1                                   & 29.8                                   \\ \cline{3-4}
\multicolumn{1}{c|}{}                       & \multicolumn{1}{c|}{}                                 & \multicolumn{1}{c|}{\multirow{3}{*}{Heterogeneous}}                                  & FedCoLLM                & 44.8                                      & 27.1                                     & 47.5                                      & 26.9                                      & 48.6                                       & 20.9                                      & 51.7                                   & 28.6                                   \\
\multicolumn{1}{c|}{}                       & \multicolumn{1}{c|}{}                                 & \multicolumn{1}{c|}{}                                                                & FedMKT                  & 46.1                                      & 26.8                                     & 46.8                                      & 27.1                                      & 49.7                                       & 28.5                                      & 53.8                                   & 30.1                                   \\
\multicolumn{1}{c|}{}                       & \multicolumn{1}{c|}{}                                 & \multicolumn{1}{c|}{}                                                                & Ours                    & \textbf{47.2}                             & \textbf{29.1}                            & \textbf{48.2}                             & \textbf{29.4}                             & \textbf{53.2}                              & \textbf{33.7}                             & \textbf{58.9}                          & \textbf{33.7}                          \\ \hline
\end{tabular}
}
\begin{tablenotes}
\footnotesize
\item The “–” symbol in the row corresponding to the standalone method indicates that this method does not belong to either the homogeneous or heterogeneous model structures setting on the devices.
\item The “–” symbol in the rows corresponding to the FedLoRA and FedAP methods indicates that these methods are not applicable for evaluating server performance, as LLMs cannot be deployed on the server under their experimental settings.
\end{tablenotes}
\end{threeparttable}
\end{table*}

\subsection{Communication Overhead Analysis}
\begin{figure}[t]
  \centering

  \begin{subfigure}[t]{\linewidth}
    \centering
    \includegraphics[width=.9\linewidth]{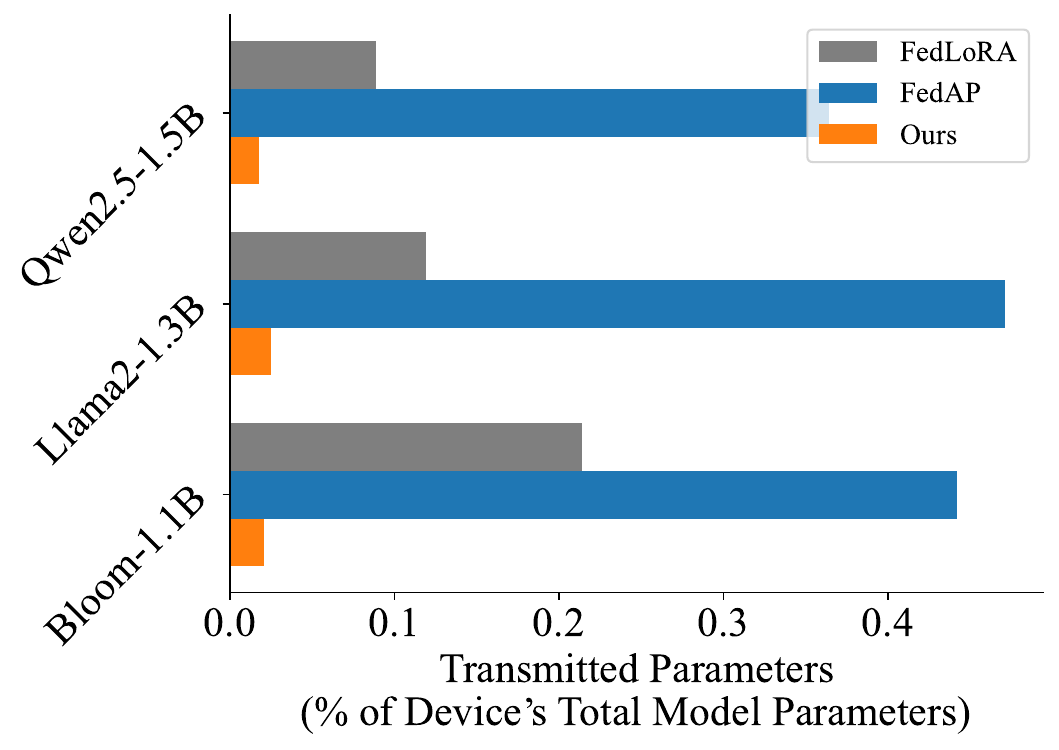}
    \caption{The setting of homogeneous model structures on the devices.}
    \label{fig:a}
  \end{subfigure}

  \vspace{0.6em} % 调整两图间距（可选）

  \begin{subfigure}[t]{\linewidth}
    \centering
    \includegraphics[width=.9\linewidth]{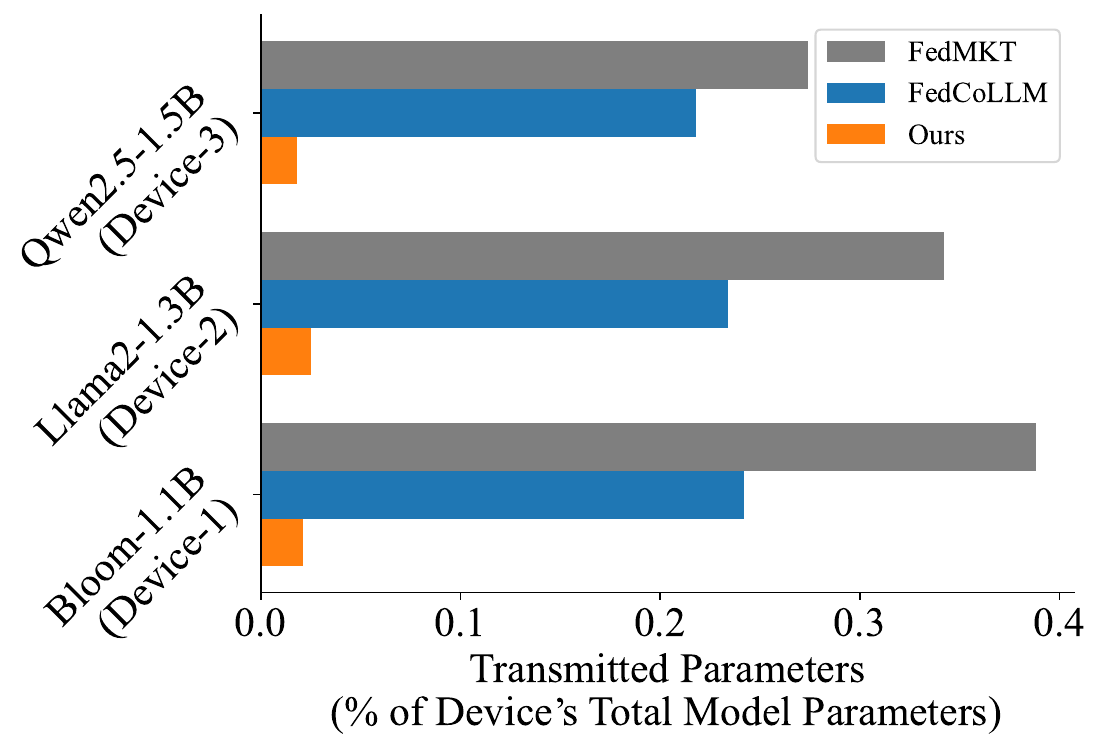}
    \caption{The setting of heterogeneous model structures on the devices.}
    \label{fig:b}
  \end{subfigure}

  \caption{Communication overheads under different settings.}
  \label{fig:stacked}
\end{figure}
To assess communication efficiency, we compare different baselines in terms of the proportion of transmitted parameters relative to the total number of model parameters on the devices.
In the setting of homogeneous model structures at the devices, all devices deploy SLMs with the same architecture and participate in the collaborative training process with an identical total number of model parameters. The communication overhead depends on the communication between the edge devices and the cloud server across different methods. As shown in Figure~\ref{fig:stacked} (a) , \pname transmits only the tunable parameters of DPM between the cloud server and devices, resulting in an average parameter transmission ratio of 0.02\%. In contrast, FedLoRa transfers low-rank adaptation matrices and FedAP transmits inserted adapter parameters, both of which require updating and communicating larger portions of the model, ranging from 0.08\% to 0.21\%, 0.36\% to 0.47\% respectively.

As illustrated in Figure~\ref{fig:stacked} (b), in FedCoLLM, each device transmits its locally trained LoRA modules to the server for model aggregation, which corresponds to a parameter transmission ratio of about 0.21\% to 0.24\%. Similarly in FedMKT, each device transmits its output logits on the dataset to the server-side LLM for transferring knowledge, the cloud server then sends its logits back to devices for SLM updates. Through bidirectional knowledge transfer, approximately 0.27\%–0.38\% of parameters are transmitted, which is substantially higher than that required by our framework. In summary, \pname achieves significantly lower communication overhead while maintaining competitive model performance.
\subsection{Ablation Study}
\begin{table}[]
\caption{Performance of \pname for ablation study}
\resizebox{1.03\linewidth}{!}{
% \usepackage{multirow}
% \usepackage{multirow}
\begin{tabular}{|lcccccccc|}
\hline
\multicolumn{9}{|c|}{\textbf{SNI}}                                                                                                                                                                                                                                                                                                                                                                                                                    \\ \hline
\multicolumn{1}{|c|}{\multirow{2}{*}{\textbf{Method}}} & \multicolumn{2}{c|}{\textbf{\begin{tabular}[c]{@{}c@{}}Bloom-1.1B\\ (Device-1)\end{tabular}}} & \multicolumn{2}{c|}{\textbf{\begin{tabular}[c]{@{}c@{}}Llama2-1.3B\\ (Device-2)\end{tabular}}} & \multicolumn{2}{c|}{\textbf{\begin{tabular}[c]{@{}c@{}}Qwen2.5-1.5B\\ (Device-3)\end{tabular}}} & \multicolumn{2}{c|}{\textbf{\begin{tabular}[c]{@{}c@{}}GPT-J-6B\\ (Server)\end{tabular}}} \\ \cline{2-9} 
\multicolumn{1}{|c|}{}                                 & \multicolumn{1}{c|}{\textbf{Rouge-L}}            & \multicolumn{1}{c|}{\textbf{EM}}           & \multicolumn{1}{c|}{\textbf{Rouge-L}}            & \multicolumn{1}{c|}{\textbf{EM}}            & \multicolumn{1}{c|}{\textbf{Rouge-L}}             & \multicolumn{1}{c|}{\textbf{EM}}            & \multicolumn{1}{c|}{\textbf{Rouge-L}}                    & \textbf{EM}                    \\ \hline
\multicolumn{1}{|l|}{\textbf{Ours}}                    & \multicolumn{1}{c|}{42.5}                        & \multicolumn{1}{c|}{31.7}                  & \multicolumn{1}{c|}{41.8}                        & \multicolumn{1}{c|}{30.2}                   & \multicolumn{1}{c|}{54.3}                         & \multicolumn{1}{c|}{39.2}                   & \multicolumn{1}{c|}{59.2}                                & 39.5                           \\ \hline
\multicolumn{1}{|l|}{\textbf{-w/o DST}}                & \multicolumn{1}{c|}{37.8}                        & \multicolumn{1}{c|}{29.1}                  & \multicolumn{1}{c|}{36.2}                        & \multicolumn{1}{c|}{27.4}                   & \multicolumn{1}{c|}{52.6}                         & \multicolumn{1}{c|}{31.7}                   & \multicolumn{1}{c|}{57.4}                                & 37.8                           \\ \hline
\multicolumn{1}{|l|}{\textbf{-w/o SAML}} & \multicolumn{1}{c|}{40.1}                        & \multicolumn{1}{c|}{29.4}                  & \multicolumn{1}{c|}{38.9}                        & \multicolumn{1}{c|}{26.7}                   & \multicolumn{1}{c|}{50.1}                         & \multicolumn{1}{c|}{32.5}                   & \multicolumn{1}{c|}{53.5}                                & 37.3                           \\ \hline\hline
\multicolumn{9}{|c|}{\textbf{MMLU}}                                                                                                                                                                                                                                                                                                                                                                                                                   \\ \hline
\multicolumn{1}{|c|}{\multirow{2}{*}{\textbf{Method}}} & \multicolumn{2}{c|}{\textbf{\begin{tabular}[c]{@{}c@{}}Bloom-1.1B\\ (Device-1)\end{tabular}}} & \multicolumn{2}{c|}{\textbf{\begin{tabular}[c]{@{}c@{}}Llama2-1.3B\\ (Device-2)\end{tabular}}} & \multicolumn{2}{c|}{\textbf{\begin{tabular}[c]{@{}c@{}}Qwen2.5-1.5B\\ (Device-3)\end{tabular}}} & \multicolumn{2}{c|}{\textbf{\begin{tabular}[c]{@{}c@{}}GPT-J-6B\\ (Server)\end{tabular}}} \\ \cline{2-9} 
\multicolumn{1}{|c|}{}                                 & \multicolumn{1}{c|}{\textbf{Rouge-L}}            & \multicolumn{1}{c|}{\textbf{EM}}           & \multicolumn{1}{c|}{\textbf{Rouge-L}}            & \multicolumn{1}{c|}{\textbf{EM}}            & \multicolumn{1}{c|}{\textbf{Rouge-L}}             & \multicolumn{1}{c|}{\textbf{EM}}            & \multicolumn{1}{c|}{\textbf{Rouge-L}}                    & \textbf{EM}                    \\ \hline
\multicolumn{1}{|l|}{\textbf{Ours}}                    & \multicolumn{1}{c|}{44.3}                        & \multicolumn{1}{c|}{31.0}                  & \multicolumn{1}{c|}{52.9}                        & \multicolumn{1}{c|}{28.5}                   & \multicolumn{1}{c|}{57.2}                         & \multicolumn{1}{c|}{39.4}                   & \multicolumn{1}{c|}{60.4}                                & 38.0                           \\ \hline
\multicolumn{1}{|l|}{\textbf{-w/o DST}}                & \multicolumn{1}{c|}{41.4}                        & \multicolumn{1}{c|}{27.9}                  & \multicolumn{1}{c|}{47.8}                        & \multicolumn{1}{c|}{24.3}                   & \multicolumn{1}{c|}{47.7}                         & \multicolumn{1}{c|}{36.1}                   & \multicolumn{1}{c|}{59.8}                                & 36.9                           \\ \hline
\multicolumn{1}{|l|}{\textbf{-w/o SAML}} & \multicolumn{1}{c|}{42.7}                        & \multicolumn{1}{c|}{28.6}                  & \multicolumn{1}{c|}{50.5}                        & \multicolumn{1}{c|}{24.9}                   & \multicolumn{1}{c|}{51.2}                         & \multicolumn{1}{c|}{31.7}                   & \multicolumn{1}{c|}{56.8}                                & 36.6                           \\ \hline
\end{tabular}
}
\label{tab:abl}
\end{table}

To validate the effectiveness of the key components in \pname, we design two variants and compare their performance against \pname on our evaluation metrics, as illustrated in Table~\ref{tab:abl} . Concretely, we investigate the contributions of the DST process and the SAML by removing each component individually and evaluating their impact on model performance across both datasets. 

\pname w/o DST variant is constructed by removing the domain-aware adapters from each Transformer layer of the DPM. In this variant, the model is trained without explicitly preserving domain-dependent knowledge, meaning that local domain-specific patterns are less effectively captured. 
The results show that \pname outperforms \pname w/o DST on both datasets, by margins of 2.90\%–4.70\%, 5.10\%–5.60\%, 1.70\%–9.50\% and 0.60\%–1.80\% across edge devices and the cloud server. This performance decline demonstrates that the DST process is crucial for mitigating the negative impact of domain heterogeneity and enabling each device model to retain its local domain knowledge.

\pname w/o SAML vairant is designed by omitting the structure-agnostic mutual learning process at the cloud server, such that the server-side LLM is not involved, and only the tunable  parameters of DPMs transmitted from each device are aggregated at the cloud server. The table displays that \pname surpasses by an average of 2.00\%, 2.70\%, 5.10\%, and 4.65\% over the w/o SAML variant. These observations confirm that SAML enables the server-side LLM to actively participate in the collaborative learning process, transferring knowledge back and forth between the cloud server and edge devices, which not only allows the cloud server to leverage insights from multiple heterogeneous client models, but also facilitates each device in improving its local SLM’s performance, enhancing both generalization and stability under resource-constrained cloud-edge system.
\section{Conclusion}
In this paper, we propose a novel co-tuning framework for collaborative training of large and small language models within cloud-edge systems, aiming to enhance model performance across edge devices and the cloud server while preserving data privacy throughout the collaborative learning process. At the start of the collaborative learning process, the DPM is distilled from the server-based LLM and distributed to all devices, where domain adapters are subsequently integrated. In each collaborative training round, every device locally fine-tunes its DPM through DST to retain domain-specific knowledge and performs SAML with its local SLM to enable bidirectional knowledge exchange. The tunable parameters of all device-side DPMs are then uploaded to the server and aggregated to update the server-based DPM. The server further conducts SAML between the updated DPM and the LLM to refine the LLM. Finally, the server redistributes the updated tunable parameters of the DPM to all devices, thereby propagating knowledge distilled from the LLM. We conduct comprehensive experiments on two datasets and compare our framework with five state-of-the-art baselines to demonstrate its effectiveness. The results show that the proposed framework consistently outperforms existing approaches under the settings of homogeneous and heterogeneous model structures at the devices throughout the collaborative learning process. In future work, we plan to investigate an optimal collaborative training process that explicitly accounts for energy costs in resource-constrained edge–cloud systems, with the goal of further improving the energy efficiency of the proposed framework.

% 云边联合体 consortium --> promising direction
% --> 还有1什么问题 --> address

\balance
\bibliographystyle{ACM-Reference-Format}
\bibliography{sample-base}

\appendix

\end{document}